\documentclass[preprint,12pt, a4paper]{elsarticle}

\usepackage{amssymb}
\usepackage[hidelinks]{hyperref}
\usepackage[nolist]{acronym}
\usepackage{epsfig}
\usepackage{orcidlink}
\journal{SoftwareX}
\biboptions{sort&compress}

\begin{document}
\renewcommand{\labelenumii}{\arabic{enumi}.\arabic{enumii}}

\begin{frontmatter}



   \title{Implementing Video Monitoring Capabilities by using hardware-based Encoders of the Raspberry Pi Zero 2 W}


   \author[1]{Thomas Ederer \orcidlink{0009-0005-1335-948X}, \href{mailto:thomas@orchideenvermehrung.at}{\includegraphics[height=7pt]{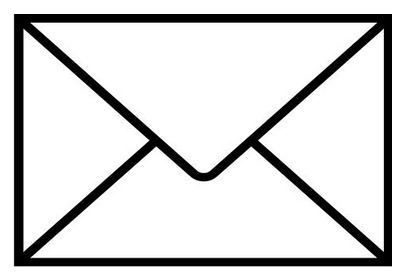}}\,}
   \author[1,2]{Igor Ivki\'c \orcidlink{0000-0003-3037-7813}, \href{mailto:i.ivkic@lancaster.ac.uk}{\includegraphics[height=7pt]{envelope.png}}\,}

   \address[1]{University of Applied Sciences Burgenland, Eisenstadt, Austria}
   \address[2]{Lancaster University, Lancaster, UK}

   \begin{abstract}
      Single-board computers, with their wide range of external interfaces, provide a cost-effective solution for studying animals and plants in their natural habitat. With the introduction of the Raspberry Pi Zero 2 W, which provides hardware-based image and video encoders, it is now possible to extend this application area to include video surveillance capabilities. This paper demonstrates a solution that offloads video stream generation from the \ac{CPU} to hardware-based encoders. The flow of data through an encoding application is described, followed by a method of accelerating image processing by reducing the number of memory copies. The paper concludes with an example use case demonstrating the application of this new feature in an underwater camera.
   \end{abstract}


\end{frontmatter}


\vspace{-15pt}
\section*{Metadata}

\vspace{-10pt}
\begin{table}[!h]
   \begin{tabular}{|l|p{6cm}|p{6cm}|}
      \hline
      \textbf{{\tiny Nr.}} & \textbf{{\tiny Code metadata description}}                               & \textbf{{\tiny Please fill in this column}}                                                                                                 \\
      \hline
      {\tiny C1}           & {\tiny Current code version}                                             & {\tiny v0.4.0}                                                                                                                              \\
      \hline
      {\tiny C2}           & {\tiny Permanent link to repository}                                     & {\tiny \href{https://github.com/tederer/octowatch-videoservice}{https://github.com/tederer/octowatch-videoservice}}                         \\
      \hline
      {\tiny C3}           & {\tiny Permanent link to Reproducible Capsule}                           & {\tiny \href{https://github.com/tederer/octowatch-videoservice/tree/v0.4.0}{https://github.com/tederer/octowatch-videoservice/tree/v0.4.0}} \\
      \hline
      {\tiny C4}           & {\tiny Legal Code License}                                               & {\tiny MIT License}                                                                                                                         \\
      \hline
      {\tiny C5}           & {\tiny Code versioning system used}                                      & {\tiny git}                                                                                                                                 \\
      \hline
      {\tiny C6}           & {\tiny Software code languages, tools, and services used}                & {\tiny C++}                                                                                                                                 \\
      \hline
      {\tiny C7}           & {\tiny Compilation requirements, operating environments \& dependencies} & {\tiny Raspbian GNU/Linux 12(bookworm)}                                                                                                     \\
      \hline
      {\tiny C8}           & {\tiny Link to developer documentation/manual}                           & {\tiny \href{https://underwater-camera-project.github.io}{https://underwater-camera-project.github.io}}                                     \\
      \hline
      {\tiny C9}           & {\tiny Support email for questions}                                      & {\tiny thomas@orchideenvermehrung.at}                                                                                                       \\
      \hline
   \end{tabular}
   \caption{Code metadata}
   \label{codeMetadata}
\end{table}


\section{Motivation and Significance}

The study of animals and plants in their natural habitat requires undisturbed observation, as this is the only way to ensure the collection of reliable data. The observational data collected helps to better understand how they function within the ecosystem and can be used to advocate for the protection of threatened environments. To achieve this goal, researchers in the field of biology \cite{jolles2021a,allan2018} are already using inexpensive single-board computers such as the Raspberry Pi \cite{mathe2024,nayyar2015} to build measurement and control systems. Previous publications \cite{almero2021,cazenave2014,coro2021,dadios2022, huang2018,jury2001,mouy2020,novy2022,phillips2019,purser2020,giddens2021,lertvilai2020,ambroz2017,baghdadi2023} demonstrate the wide range of applications for the \textit{Raspberry Pi}. These projects have in common the connection of a camera module to a single-board computer. Software executing on the \ac{CPU} of the computer is responsible for converting the camera's raw images into a video stream. However, this type of video generation consumes computing time of the \ac{CPU}, which cannot be used for other tasks. The following figure shows a comparison of software- and hardware-based \ac{JPEG} encoders.

\begin{figure}[!h]
   \centering
   \includegraphics[width=0.6\textwidth]{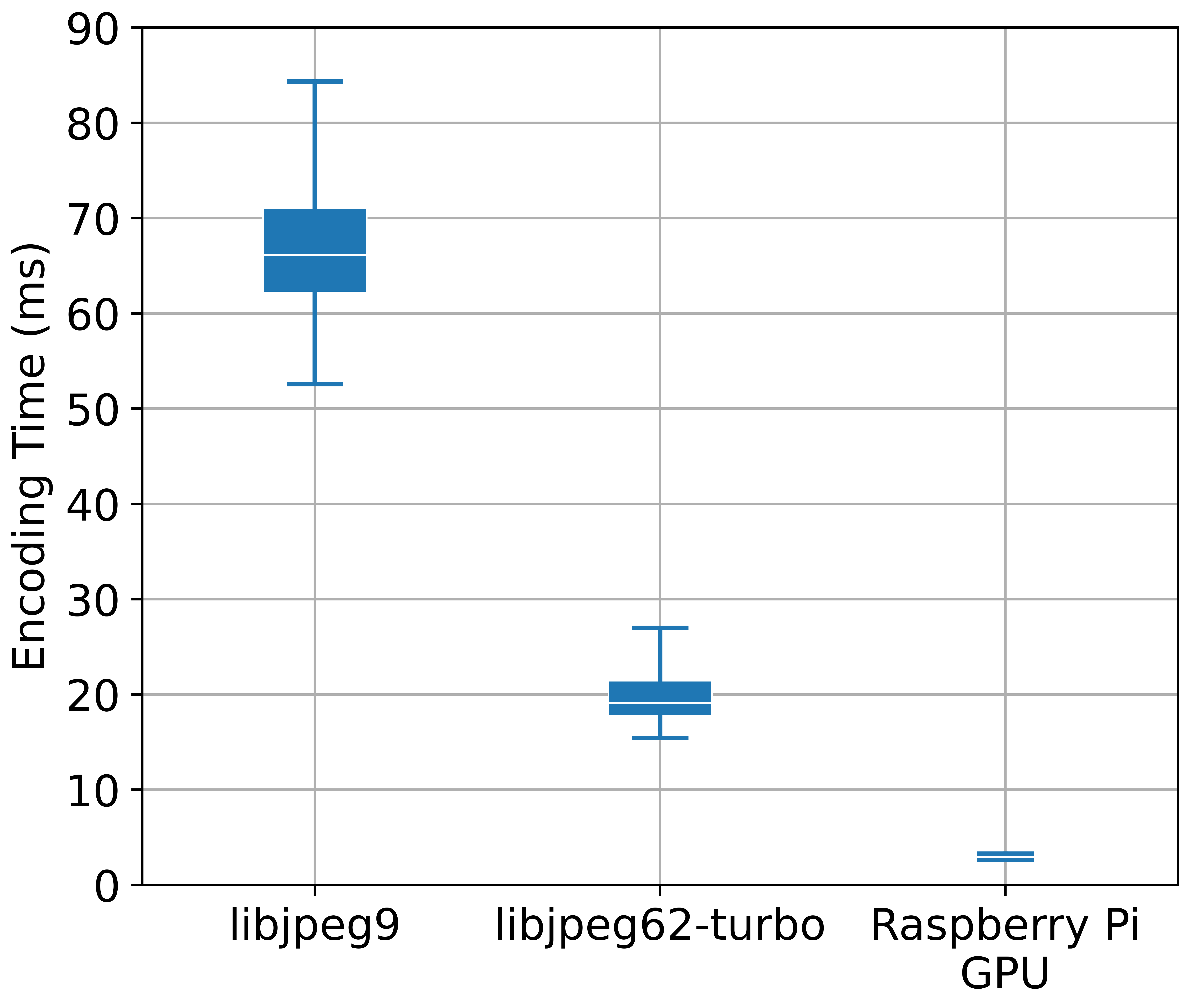}
   \caption{Comparing the Encoding Time required for Converting Raw Camera Images (YUV420, 800 x 600 pixels) to \ac{JPEG} Format using various Encoding Methods.}
   \label{fig:jpegEncodingPerformance}
\end{figure}

The performance of the encoding was evaluated by measuring the time taken to encode 1,000 images. Before using the images, they were converted to the output format (YUV420) of the camera module, and then each image was encoded 100 times. The time required for this process was measured, and the results demonstrated that the speed of \textit{libjpeg9} is not sufficient to produce a video stream with 30 frames per second. Using \textit{libjpeg62-turbo} is a viable option; however, it occupies one \ac{CPU} core almost entirely. The optimal solution is to perform the encoding of the images in hardware, as this is fast (approximately 4 milliseconds per image) and reduces the load on the \ac{CPU}.

In our previous publication \cite{ederer2024a}, we demonstrated a way to overcome this limitation by using the hardware-based image and video encoders of the \textit{Raspberry Pi Zero 2 W} for video stream generation. An underwater camera for remote monitoring of octopuses in the northern Adriatic Sea was developed to generate two video streams: one for local recording and another integrated into a web page for remote access via the Internet. This eliminates the necessity of observers being on-site, thereby reducing the on-site presence and assisting research projects in undertaking their observations in a more cost-effective manner. The following figure shows the housing (made of water pipes) of the underwater camera and its components mounted in an adaptable installation framework:

\vspace{10px}
\begin{figure}[!h]
   \centering
   \includegraphics[width=1.0\textwidth]{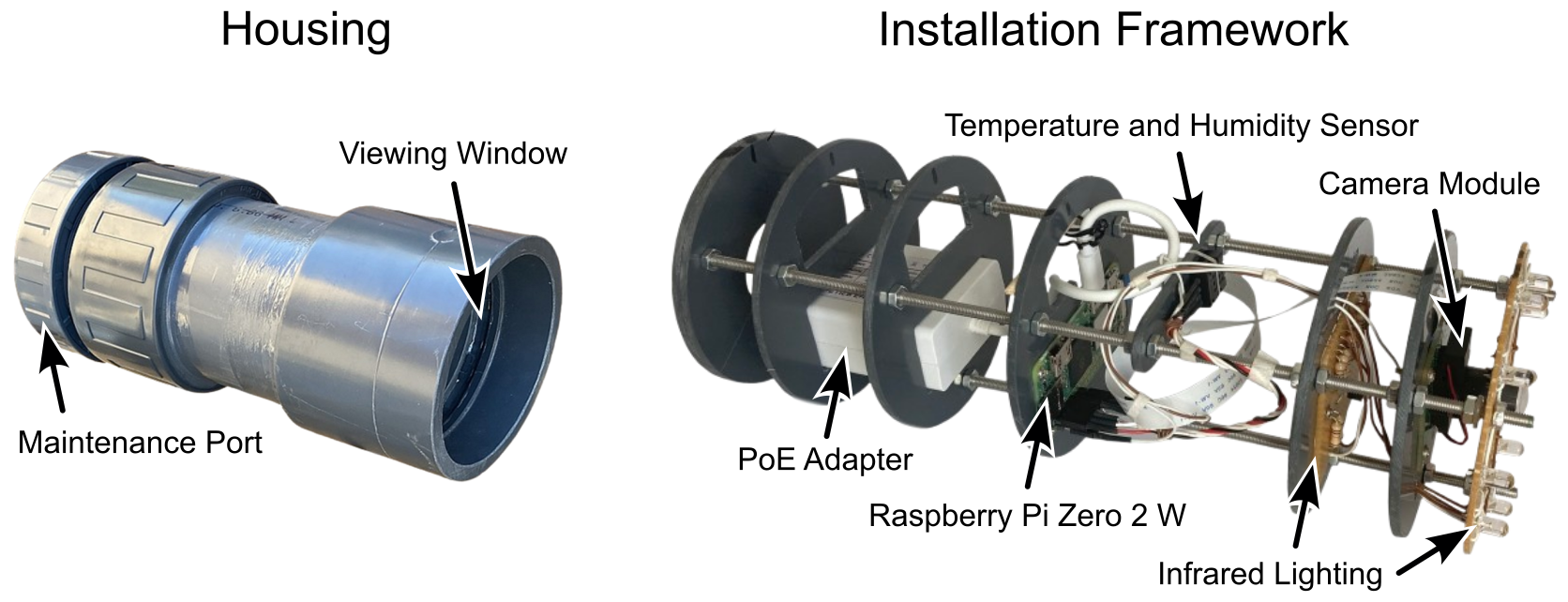}\caption{Underwater Camera using the hardware-based Image and Video Encoders of the Raspberry Pi Zero 2 W for Video Stream Generation.}
   \label{fig:camera}
\end{figure}

This paper is a continuation of \cite{ederer2024a} and shows how to use the hardware-based video and image encoders of the \textit{Raspberry Pi Zero 2 W} to integrate video surveillance capabilities into existing measurement and control tasks. First, we present the overall software architecture of the video service and explain how the data flow is used to generate two video streams from a sequence of raw images.
Next, we discuss the effectiveness of the underlying framework in minimizing image copying. Finally, we show how a combination of implemented software and freely available hardware components can be used to provide two independent video streams in an underwater camera.


\section{Software description}

The purpose of the Video Service is to convert the images from the camera module into two independent video streams and make them available over the network. At startup, the service configures the camera module and then waits for images from the scene in front of the camera. Each image received is passed to the hardware-based encoders for conversion to H.264 and \ac{MPJPEG} formats, which generate a sequence of \ac{NAL} packets (for H.264) and \ac{JPEG} images (for \ac{MPJPEG}). Before the packets are sent, each \ac{JPEG} image must be augmented with additional data so that it can be processed by the receiver as a \ac{MPJPEG} video stream. The converted data can be consumed independently of each other via separate network ports.


\subsection{Software architecture}

\begin{figure}[!h]
   \centering
   \includegraphics[width=0.7\textwidth]{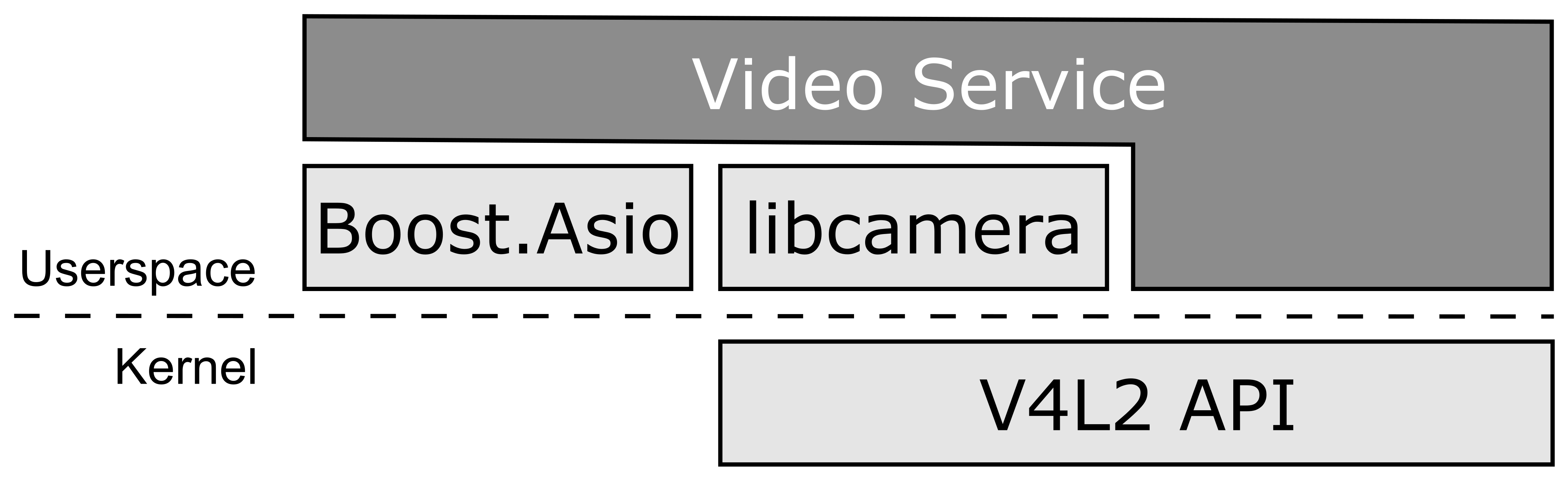}
   \caption{Software stack of the Video Service, consisting of existing Components (light grey) and those to be implemented (dark grey).}
   \label{fig:videoServiceStack}
\end{figure}

The software is designed as a Linux service implemented in the programming language C++ on top of the libraries \textit{libcamera} (v0.4.0), \textit{Boost.Asio} (1.82.0) and the \ac{V4L2} \ac{API} (linux-libc-dev:6.1.140-1). The object-oriented interface of \textit{libcamera} reduces the time taken to integrate camera modules into applications by hiding device-specific implementation details. Figure \ref{fig:videoServiceStack} illustrates the software stack of the Video Service consisting of existing components (light grey) and those to be implemented (dark grey).

As hardware-based encoders are not yet supported by the \textit{libcamera} library, the underlying \ac{V4L2} \ac{API} had to be used for this part of the implementation. The network communication components are based on \textit{Boost.Asio}, which is a cross-platform C++ library for network and low-level I/O programming. It is possible to modify camera settings during runtime by utilizing the Remote Control Interface of the service.
The core of the Video Service is a game loop \cite{valente2005}, which uses a number of components. At the consuming end the service receives two images in different resolutions from the camera module and transfers them to the corresponding encoder. This offloads image scaling from the \ac{CPU} to the camera module. Images with a high resolution (1920 x 1080 pixels) are used to generate the H.264 video stream \cite{itut_h264}, while those with a lower resolution (800 x 600 pixels) are used to generate the \ac{MPJPEG} video stream. Furthermore, the multipart streamer component adds metadata \cite{borenstein1992} to each JPEG image to enable web browsers to render the \ac{MPJPEG} stream as part of a web page. In order to facilitate the delivery of the two video streams via the network, two TCP servers are used, each accepting connections on a distinct port. The following figure illustrates the data flow in the Video Service:

\begin{figure}[!h]
   \centering
   \includegraphics[width=\textwidth]{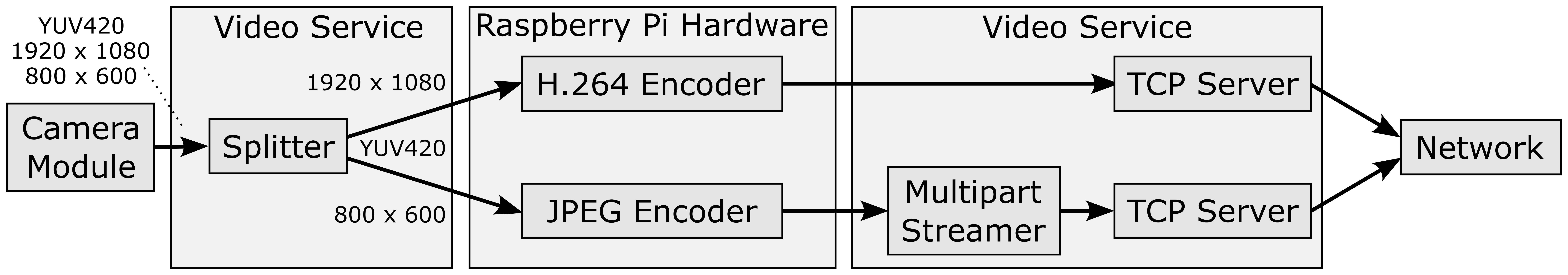}
   \caption{Data flow in the Video Service used to Produce two Video Streams from a Sequence of Raw Images.}
   \label{fig:videoServiceDataFlow}
\end{figure}

To convert the sequence of images (provided by the camera module) into a video stream, each image has to be processed by different components of the Video Service. In the worst case, images have to be copied when they are provided to, for example, the encoder. Such copy operations take a long time and have a large impact on the maximum throughput. To reduce copying to a minimum, \ac{V4L2} uses the \textit{DMABUF} method. When a new image is downloaded from the camera module, the consumer of the image receives two integer numbers, a file descriptor and a memory offset. With these two values it is possible to map the memory of the image to the memory of the current process without copying data. This reduces the number of copies to one, as the network interface is the only one that requires a copy of the data due to the asynchronous nature of the data transfer.


\subsection{Software functionalities}

The video service presented can be used to integrate video surveillance capabilities into control and measurement systems using a low-cost \textit{Raspberry Pi Zero 2 W} and a camera module. By using the service, the camera module gets configured to provide two images with different resolutions to generate two video streams: one in H.264 format and another in \ac{MPJPEG} format. There is also an interface to modify the properties of the camera module.

\subsubsection{H.264 video stream}

The video stream is provided by the service via network port 8888, with a resolution of 1920 x 1080 pixels. Using H.264 encoding with its ability to compress data over multiple frames in the video stream ensures a bandwidth requirement of 10 MBit/s. The encoded video data is transmitted using \ac{NAL} packets, which can be stored locally or on a server in the cloud for later playback and processing. Real-time playback can be facilitated by using the free and open source cross-platform multimedia player \textit{VLC}, for example. Free tools such as \textit{FFmpeg} can be used for recording the video stream.

\subsubsection{\ac{MPJPEG} video stream}

Network port 8887 is used by the service for the transmission of the stream, with a resolution of 800 x 600 pixels. \ac{MPJPEG} encoding enables the integration of the stream into web pages, and by adjusting the image compression rate to its maximum, a bandwidth requirement of 2.5 MBit/s can be achieved. The combination of H.264 for local recording and \ac{MPJPEG} for internet transmission can reduce the on-site presence, thereby assisting research projects in conducting their observations in a more cost-effective manner. Environment variables can be used to configure the compression rate and the type (\ac{GPU} or \ac{CPU}) of the JPEG encoder used.

\subsubsection{Required Network Bandwidth}

\begin{figure}[!h]
   \centering
   \includegraphics[width=1\textwidth]{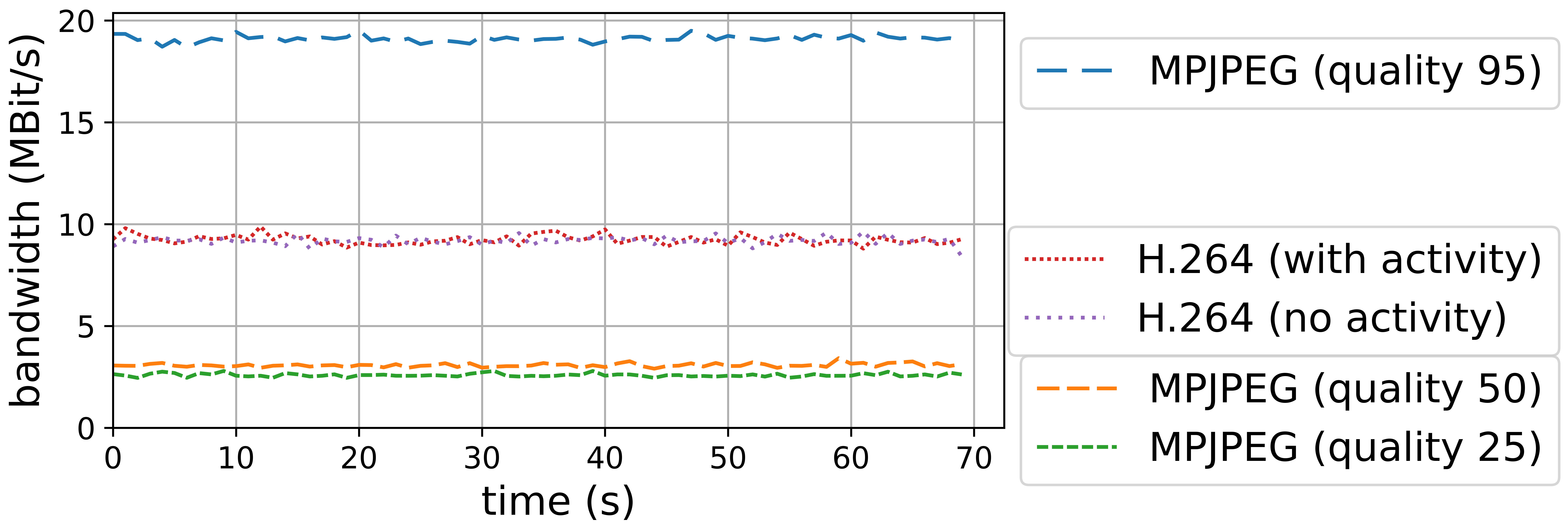}\caption{Required Network Bandwidth depending on Encoding and Image Quality.}
   \label{fig:bandwidth}
\end{figure}

In order to measure the required network bandwidth (Figure \ref{fig:bandwidth}) for the transmission of the described video streams, three scenarios were selected for the MPJPEG video stream, covering the range of supported image quality values (0 – 95). Two scenarios were used to evaluate the network bandwidth of the H.264 video stream: (1) with no activity in front of the camera, and (2) with activity in the recorded scene.

\section{Illustrative example}

In the context of the \textit{Octopus Intelligence} project of \href{https://mare-mundi.org/}{\textbf{Mare Mundi}}, an underwater camera was constructed for the observation of octopuses in the northern Adriatic Sea \cite{ederer2024a}. The video streams were generated using a \textit{Raspberry Pi Zero 2 W}, a camera module and the Video Service presented in this paper. Given the high compression rate, the H.264 format was used to record the camera output on-site for later processing. In order to facilitate real-time observation, additional services had to be developed and are available for free at the \href{https://underwater-camera-project.github.io/}{\textbf{Underwater Camera Project}}. A web server was used to provide a web page \cite{mikowski2014} containing the \ac{MPJPEG} video stream available from any location with internet access, as well as controls for modifying the camera settings (e.g. brightness). The following figure illustrates the architecture of the video-related components of the underwater camera:

\begin{figure}[!h]
   \centering
   \includegraphics[width=\textwidth]{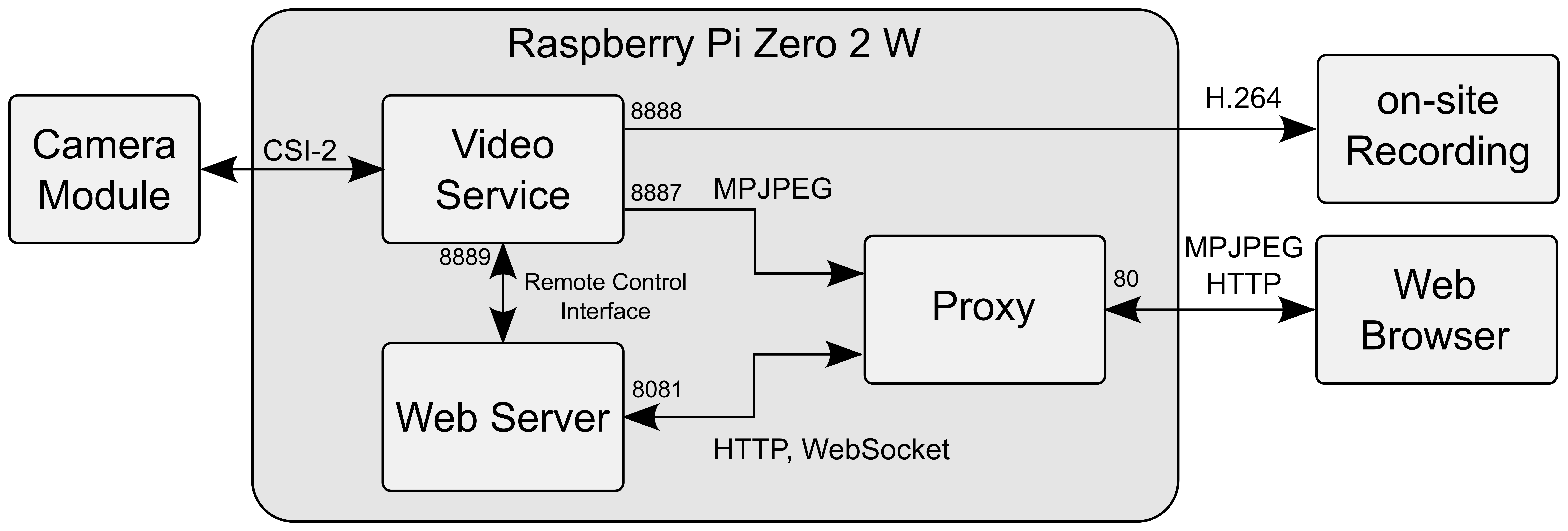}\caption{Video Related Part of the Architecture used to Monitor Octopuses in the northern Adriatic Sea.}
   \label{fig:applicationArchitecture}
\end{figure}

A proxy is required to make the content of the Video Service and the web server available on the same port. Direct access to the web server is technically possible, but it makes it impossible for web browsers to determine whether the content on the website is from the same source as the video. As a result, the video download is blocked by the browser. A proxy (like \textit{nginx}) or application gateway hides both services behind a single interface to avoid requests getting blocked. All those services were configured to run as Linux services. This enables the operating system of the \textit{Raspberry Pi} to start them at system boot and to collect their log output in the system journal. To reduce the risk of getting attacked, the firewall was configured to drop all incoming connections to port which are different to those of the H.264 video stream and the proxy.

%
%
%
%
%
%

\section{Impact}

The Video Service presented was developed as part of the implementation of a low-cost underwater camera used by \href{https://mare-mundi.org/}{\textbf{Mare Mundi}} to study the intelligence of octopuses in the northern Adriatic Sea \cite{ederer2024a}. The findings demonstrate that the hardware-based video and image encoders of the Raspberry Pi Zero 2 W possess sufficient capabilities to concurrently generate two video streams (H.264 and \ac{MPJPEG}) with differing transmission bandwidth requirements.
This facilitates the transmission of the video stream over the internet and its local recording. In contrast, commercially available camera systems generally provide a single video stream, without the possibility of customization, such as adding additional sensors or actors. The proposed solution overcomes these limitations by leveraging the \textit{Raspberry Pi's} ability to provide a variety of external interfaces for connecting devices, and its software can be modified and extended by anyone. From a financial perspective, the transmission of video content via the internet, in conjunction with local recording, has the potential to reduce the necessity for on-site researchers, thereby leading to cost reductions while maintaining the capacity to conduct observations. The local recording of video output is another potential cost-saving measure, as it allows users to access videos when they have time to do so.

In the context of future projects, the potential exists for detecting activity in the video and providing users with the corresponding sequences from the high quality local recording for later analysis. The elimination of time-consuming viewing of the recording further reduces project costs.

\section{Conclusions}

The use of video surveillance presents certain challenges for research projects, as the acquisition of cameras is expensive and the devices cannot be extended with sensors and actuators. Low-cost single-board computers such as the \textit{Raspberry Pi} have demonstrated their efficacy for measurement and control tasks in numerous publications. With the \textit{Raspberry Pi Zero 2 W's} new hardware-based video and image encoders, it is now possible to produce high quality video at low cost. In this paper, we use a practical example to show how the efficient use of encoders has turned the Raspberry Pi into an underwater camera that generates two video streams.
One stream (H.264) is used for on-site recording, while the other video stream (\ac{MPJPEG}) is embedded in a website that can be accessed via the Internet. Using these two video streams allows researchers to observe the underwater environment without disturbing it with their presence, also eliminating the need to be present on site and reducing the costs.



\bibliographystyle{elsarticle-num}
\bibliography{references}







\begin{acronym}[MPJPEG] 
   \acro{API}[API]{Application Programming Interface}
   \acro{COTS}[COTS]{Commercial Off The Shelf}
   \acro{CSI-2}[CSI-2]{Camera Serial Interface 2}
   \acro{CPU}[CPU]{Central Processing Unit}
   \acro{GPU}[GPU]{Graphics Processing Unit}
   \acro{HTML}[HTML]{Hyper Text Markup Language}
   \acro{JPEG}[JPEG]{Joint Photographic Experts Group}
   \acro{MPJPEG}[MPJPEG]{Multipart JPEG}
   \acro{NAL}[NAL]{Network Abstraction Layer}
   \acro{SIMD}[SIMD]{Single Instruction Multiple Data}
   \acro{SoC}[SoC]{System-on-a-Chip}
   \acro{TCP}[TCP]{Transmission Control Protocol}
   \acro{V4L2}[V4L2]{Video for Linux API Version 2}
\end{acronym}

\end{document}